\documentclass[12pt,a4paper,dvips]{article}
\usepackage{a4p}
\usepackage{epsfig,subfigure}
\usepackage{cite,mcite}
\usepackage{graphicx}
\usepackage{physics }
\usepackage{l3_title,ifthen}
\journalname{Phys. Lett. B}
\date{April 12, 2000}
\preprint{2000-054}
\newlength{\capindent}
\newlength{\capwidth}
\newlength{\figwidth}
\newcommand{\icaption}[2][!*!,!]{\hspace*{\capindent}%
  \begin{minipage}{\capwidth}
    \ifthenelse{\equal{#1}{!*!,!}}%
      {\caption{#2}}%
      {\caption[#1]{#2}}
  \end{minipage}}
\newcommand{\pho}{\phantom{0}}
\newcommand{\BB}{\ensuremath{\mathrm{b\bar{b}}}}
\newcommand{\COM}{\ensuremath{\mbox{centre--of--mass}}}
\newcommand{\RBB}{\ensuremath{R_{\mathrm{b}}}}
\newcommand{\RCC}{\ensuremath{R_{\mathrm{c}}}}
\newcommand{\EE} {\ensuremath{\mathrm{e^+e^-}}}

\newcommand{\QQ} {\ensuremath{\rm q\bar{q}}}
\newcommand{\QQG}{\ensuremath{\QQ(\gamma)}}
\newcommand{\AFB}{\ensuremath{A_{\mathrm{fb}}}}
\newcommand{\AFBB}{\ensuremath{\AFB^{\mathrm{b}}}}
\newcommand{\RS}{\ensuremath{\sqrt{s}}}
\newcommand{\WW} {\ensuremath{\rm W^+W^-}}
\newcommand{\WWG}{\ensuremath{\WW(\gamma)}}
\newcommand{\HA} {\ensuremath{\rm hadrons}}
\newcommand{\HAG}{\ensuremath{\HA(\gamma)}}
\newcommand{\EHAG}{\ensuremath{\rm \EE\rightarrow\HAG}}

\newcommand{\ZE} {\ensuremath{\rm Zee}}

\newcommand{\TT} {\ensuremath{\rm \tau^+\tau^-}}
\newcommand{\TTG}{\ensuremath{\TT(\gamma)}}

\newcommand{\EWWG}{\ensuremath{\rm \EE\rightarrow\WWG}}
\newcommand{\BOBO} { $\mathrm{B}^{0}\bar{\mathrm{B}}^{0}$ }
\newcommand{\AFBBm}{\ensuremath{A_{\mathrm{fb}}^{\mathrm{b,\mu}}}}
\newcommand{\AFBBe}{\ensuremath{A_{\mathrm{fb}}^{\mathrm{b,e}}}}
\newcommand{\IL}{\ensuremath{{\cal L}}}
\newcommand{\IVPB}{\ensuremath{\mathrm{pb^{-1}}}}
\newcommand{\AFBC}{\ensuremath{\AFB^{\mathrm{c}}}}
\newcommand{\SM}{Standard Model}
\newcommand{\Do}  {\ensuremath{\mathrm{D^0}}}
\newcommand{\Dp}  {\ensuremath{\mathrm{D^+}}}
\newcommand{\Ds}  {\ensuremath{\mathrm{D_s}}}
\newcommand{\Lc}  {\ensuremath{\mathrm{\Lambda_c}}}
\begin{document}
\begin{titlepage}
\title{\boldmath Measurements of the \BB{} production cross section and
  forward--backward asymmetry at \COM{} energies above the Z pole at LEP}
\author{The L3 Collaboration}
%
%

\begin{abstract}
  The measurements of $\RBB = \sigma(\EE \to \BB)/\sigma(\EE
  \to \QQ)$ and of the b quark forward--backward charge asymmetry,
  \AFBB{}, at centre--of--mass energies above the Z pole are
  described. The measurement of \RBB{} is performed at \RS{}
  between 130 and 189 \GeV{} using a b--tagging method that
  exploits the relatively large decay length of
  b--hadrons. The measurement of \AFBB{} is performed using the large
  statistics event sample collected at $\RS=189\GeV$ with
  a lepton--tag analysis based on the selection of prompt muons and
  electrons. The results at $\RS=189\GeV$ are:
\begin{eqnarray*}
\RBB  &=& 0.163 \pm 0.013~(stat.) \pm 0.005~(syst.),\\
\AFBB &=& 0.61 \pm 0.18~(stat.) \pm 0.09~(syst.).
\end{eqnarray*}
\end{abstract}
\centerline{\it{Dedicated to the memory of Prof. Bianca Monteleoni}}
\vspace*{-10mm}
\submitted

\end{titlepage}
\section*{Introduction}
The ratio $\RBB = \sigma(\EE \to \BB)/\sigma(\EE \to \QQ)$ and the
bottom quark forward--backward asymmetry \AFBB{} are important
parameters in precision studies of the Standard Model \cite{SM}
and are sensitive probes to new physics.
Their values have been measured very
precisely at the Z pole
\cite{RBBAFBL1A,RBBAFBL1D,RBBAFBL1L,RBBAFBL1O,RBBAFBSLD}. 
The measurements of these
two quantities at centre--of--mass energies above the Z peak provide
further tests of the Standard Model and put additional
constraints on new physics, such as contact
interactions \cite{THEORY_CI}.

In this paper the measurements of \RBB{} at centre--of--mass energies
between 130 and 189 \GeV{} and of \AFBB{} at $\RS=189\GeV$ obtained 
with the L3 detector \cite{L3DET} at LEP are
described. 
The main features that distinguish the 
production of $\mathrm{b\bar{b}}$ pairs from lighter quark production
are the long lifetime and hard fragmentation of b--flavoured hadrons
and the large lepton momentum in semileptonic decays.
The measurement of \RBB{} is based on a b--tagging method using lifetime
information which uses data taken with the
L3 Silicon Microvertex Detector (SMD) \cite{L3-SMD}. The measurement of \AFBB{}
exploits the characteristic semileptonic decays of b--hadrons and
relies on the good lepton identification and lepton energy resolution of the
L3 detector. It requires leptons with high momentum along and 
transverse to the direction of the associated jet, caused by the hard 
fragmentation and high mass of the decaying hadron.
Similar measurements have been published by other LEP
collaborations \cite{RBBAFBLEP2}.

\section*{Event selection}

The data analysed in this paper are those collected by L3 from 1995 to
1998 at centre--of--mass energies between 130 \GeV{} and 189 \GeV{}.
As the integrated luminosity at 130 \GeV{} and 136 \GeV{} is small,
the corresponding data are combined and a luminosity weighted average
centre--of--mass energy of 133.2 \GeV{} is used.  The centre--of--mass
energies and the corresponding integrated luminosities used in the
analysis are summarised in Table~\ref{tab:enelumi}.

The measurements of \RBB{} and \AFBB{} are performed using a sample of
$\EE \to {\rm Z^\ast}/\gamma^\ast \to \QQG$ events selected with criteria
similar to those used for the measurement of the $\EE\to{\rm hadrons}$
cross section \cite{LINESHAPE189}. The signal events are required to
have an effective centre--of--mass energy
$\sqrt{s^\prime}>0.85\RS$.  This requirement rejects a large fraction
of events with initial state radiation. Mis-reconstruction of the
effective \COM{} energy induces a migration of radiative events to the
kinematic region allowed by the cut on $\sqrt{s^\prime}$. This is
taken into account in the analysis as an additional background,
called radiative background. No correction is applied to the
final results to take into account the interference
between initial-- and final--state radiative corrections.

To further reject radiative events, the visible energy, $E_{\rm vis}$, is
required to be greater than $0.6\RS$ and the longitudinal energy
imbalance must be less than $0.25\,E_{\rm vis}$.  The events
are selected in a fiducial region defined by $|\cos\theta\,|<0.85$,
where $\theta$ is the polar angle of the thrust axis. This cut
ensures good track quality and efficient lepton identification.  For
\COM{} energies larger than the W--pair production threshold,
additional cuts are needed to reject the \WW{} background.  To reject
hadronic W decays, the variable $y_{34}$ is
required to be less than 0.011, where $y_{34}$ is the Durham jet
finding algorithm parameter \cite{DURHAM} for which the transition from three
to four jets occurs.  As the \WW{} background increases for higher
\COM{} energies, the maximum allowed value of $y_{34}$ is lowered to
0.008 at $\RS=183\GeV$ and to 0.006 at 189 \GeV{}. Furthermore, to
reject semileptonic W decays, the transverse energy imbalance must be
smaller than $0.4\,E_{\rm vis}$.  For the \AFBB{} analysis the
presence of an identified electron or muon with an energy smaller than
40 \GeV{} is required.

Efficiency and background studies are performed for each
centre--of--mass energy using the following event generators:
PYTHIA \cite{PYTHIA} (\EHAG, \ZE), KORALZ \cite{KORALZ} ($\TTG$),
PHOJET \cite{PHOJET} (hadronic two--photon collisions),
KORALW \cite{KORALW} (\EWWG) and EXCALIBUR \cite{EXCALIBUR} ($\EE\to
{\rm q \bar q q^{\prime}{\bar q}^{\prime}}$), which includes the ZZ
production diagrams.

\section*{\boldmath Measurement of \RBB}

The measurement of \RBB{} is based on a 
b--tagging algorithm that exploits the relatively large decay
length of b--hadrons\cite{HIGGSBTAG}. The confidence level, $C_N$, that a set of
$N$ tracks originated from the primary vertex, is constructed using the 
decay length significance $L/\sigma_L$ of each track. 
First the crossing point of
each track with the closest jet is determined in both the $r\phi$ and
$sz$ projections. Then the signed distances ($d_{r\phi}$, $d_{sz}$) 
between these crossing points
and the reconstructed primary event vertex are projected onto the jet
axis to determine the decay length, $L$. If the probability that both
$r\phi$ and $sz$ measurements are compatible exceeds 5\%, then the two
are combined. Otherwise, only the $r\phi$ projection is used.

Small biases in $d_{r\phi}$ and $d_{sz}$ are removed by recalibrating
their mean values as functions of the azimuthal and polar angles of
the tracks. 
The errors on $d_{r\phi}$ and $d_{sz}$ are parametrised to take into
account the
contributions coming from the track fit, the primary vertex position
measurement and the effect of multiple scattering. The estimated
errors on $d_{r\phi}$ and $d_{sz}$ are
rescaled by comparing their values with those observed in a sample
of tracks with negative decay length. The scale factors are found to
be within 10\% of unity.
The tracking resolution depends critically on the pattern of SMD hits
associated to the tracks. Therefore the tracks are grouped into
different classes according to the vertex detector hit pattern and for
each class different resolution parameters are considered.
The study of the tracking resolution described
above is performed using samples of hadronic Z decays corresponding
to an integrated luminosity of about $2.5\, {\rm pb^{-1}}$ each, collected
each year during calibration runs at the Z peak.

\begin{figure}
  \vspace*{-2cm}
  \begin{center}
    {\epsfig{figure=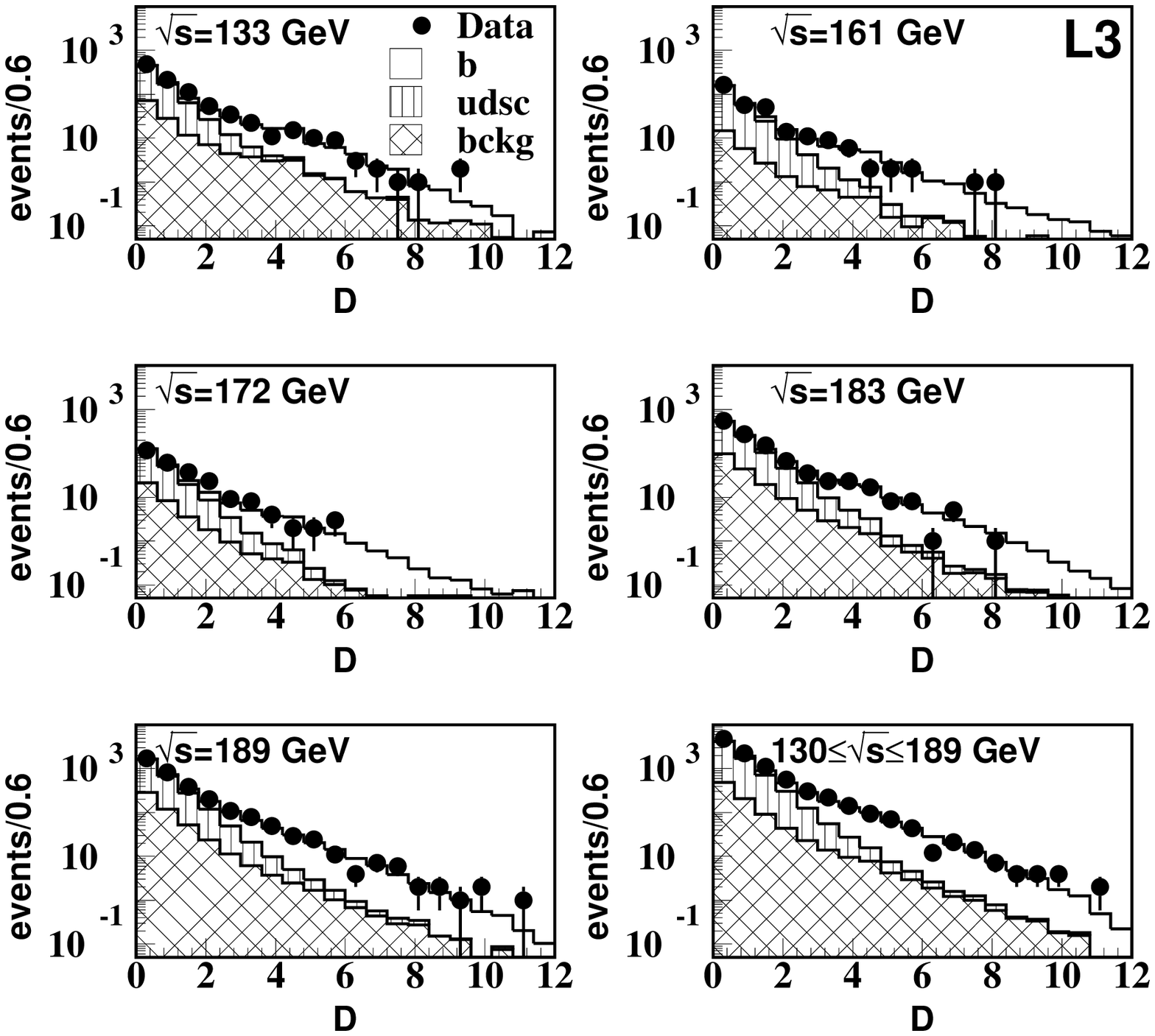,width=\textwidth}}
    \icaption{Distributions of the b--tagging discriminant for the
      selected events at the different centre--of--mass energies. The
      points are the data and the solid lines represent the total Monte Carlo
      expectation. The open histograms show the signal
      contribution, the vertically hatched histograms represent the
      c and light quark background from \QQ{} events and  
      the cross--hatched histograms represent the sum of non--\QQ\
      and radiative background, as
      indicated in the legend of the upper left figure. 
      In the bottom right plot, the distribution for the total selected
      sample is shown.
    \label{fig:discrim}}
  \end{center}
\end{figure}

The confidence level, $C_N$, is calculated by taking
into account the fraction of tracks with positive decay length,
\begin{equation}
C_{N}=\frac{\Pi}{2^N}
\sum_{i=0}^{N-1} \sum_{j=i+1}^N \left( \begin{array}{c} N\\j \end{array} \right)
\frac{(-\log\Pi)^i}{i!}\;,   \;\;\;\; \Pi = \prod_{k=1}^{N_+} P_k(L/\sigma_L),
\end{equation}
where $N_+$ is the number of tracks with positive decay length.
The probability that a track originated from the primary vertex,
$P(L/\sigma_L)$, is obtained by fitting the distribution of the decay
length significance for tracks with negative decay length. The shape
of this distribution is due to detector resolution effects and is
modelled by a resolution function parametrised as the sum of two
gaussians and an exponential tail. The distributions of the
discriminant variable $D=-\log_{10}(C_{N})$ obtained at the different
centre--of--mass energies are shown in Figure~\ref{fig:discrim}.

The measurement of \RBB{} is performed using an event--tag method by
counting the events containing b quarks. The b--events are selected by
applying a cut on the discriminant where the cut position is chosen so
that the expected statistical error on the cross section
$\sigma(\EE\to\BB)$ is minimised.
Given the number $N\rm{_t^{obs}}$ of tagged
events and $N\rm{_t^{bkg}}$ of tagged background events, \RBB{} is 
derived from the formula:
\begin{equation}
\RBB = \frac{1}{\varepsilon{\rm_b - \varepsilon_{uds}}}\left[\frac{N{\rm_t^{obs}}-
N{\rm_t^{bkg}}}{N{\rm^{obs}}-N{\rm^{bkg}}} -
    \varepsilon \rm{_c}\RCC - \varepsilon_{\rm{uds}}(1-\RCC)\right],
\label{eq:rbbdef}
\end{equation}
where $N\rm{^{obs}}$ and $N\rm{^{bkg}}$ are respectively the total number of
selected \QQ{} events in data and the total number of background events
from Monte Carlo; $\varepsilon_{\rm b}$, $\varepsilon_{\rm{c}}$
and $\varepsilon_{\rm{uds}}$ are the tagging efficiencies for b, c and
light quarks, respectively, and are estimated from Monte Carlo.

The
value of the ${\rm c\bar c}$ production cross section, \RCC{}, is taken
from the Standard Model prediction calculated by
ZFITTER \cite{ZFITTER}.
The dependence of \RBB{} on \RCC{} can be parametrised as 
\begin{equation}
\Delta\RBB = a(\RCC)\Delta\RCC
\label{eq:rccdep}
\end{equation}
where $\Delta\RCC = \RCC-\RCC^{\rm SM}$. 
The values of $\RCC^{\rm SM}$ used in the measurement and the corresponding
coefficients $a(\RCC)$ are shown in Table~\ref{tab:rccdep}.

The background includes the contribution from non--\QQ{} events and the
contamination of radiative hadronic events. The radiative background
decreases from 16\% at 133 \GeV{} to 4\% at 189 \GeV{}. 
The background due to
processes other than $\EE\to\QQG$ is dominated by hadronic W decays for
$\RS\geq 161\GeV$ and varies from 1\% at $\RS=161\GeV$ to 11\% at
189 \GeV{}. The background from $\EE\to
{\rm q \bar q q^{\prime}{\bar q}^{\prime}}$ events is at the level of 1\% at
189 \GeV{} and negligible at lower \COM{} energies. The contribution
from other processes is negligible. 

The number of selected and tagged events and the background
contamination both in the total and in the tagged sample for each
centre--of--mass energy are shown in
Table~\ref{tab:rbbresults}. The tagging efficiencies are also quoted
along with their statistical errors. The b tagging
efficiency decreases slightly with increasing \COM{} energy due to the
larger non--\QQ{} background contribution. The measured values of
\RBB{} are shown in Table~\ref{tab:rbbresults} together with their
statistical errors.

\begin{figure}[ht]
  \begin{center}
    {\epsfig{figure=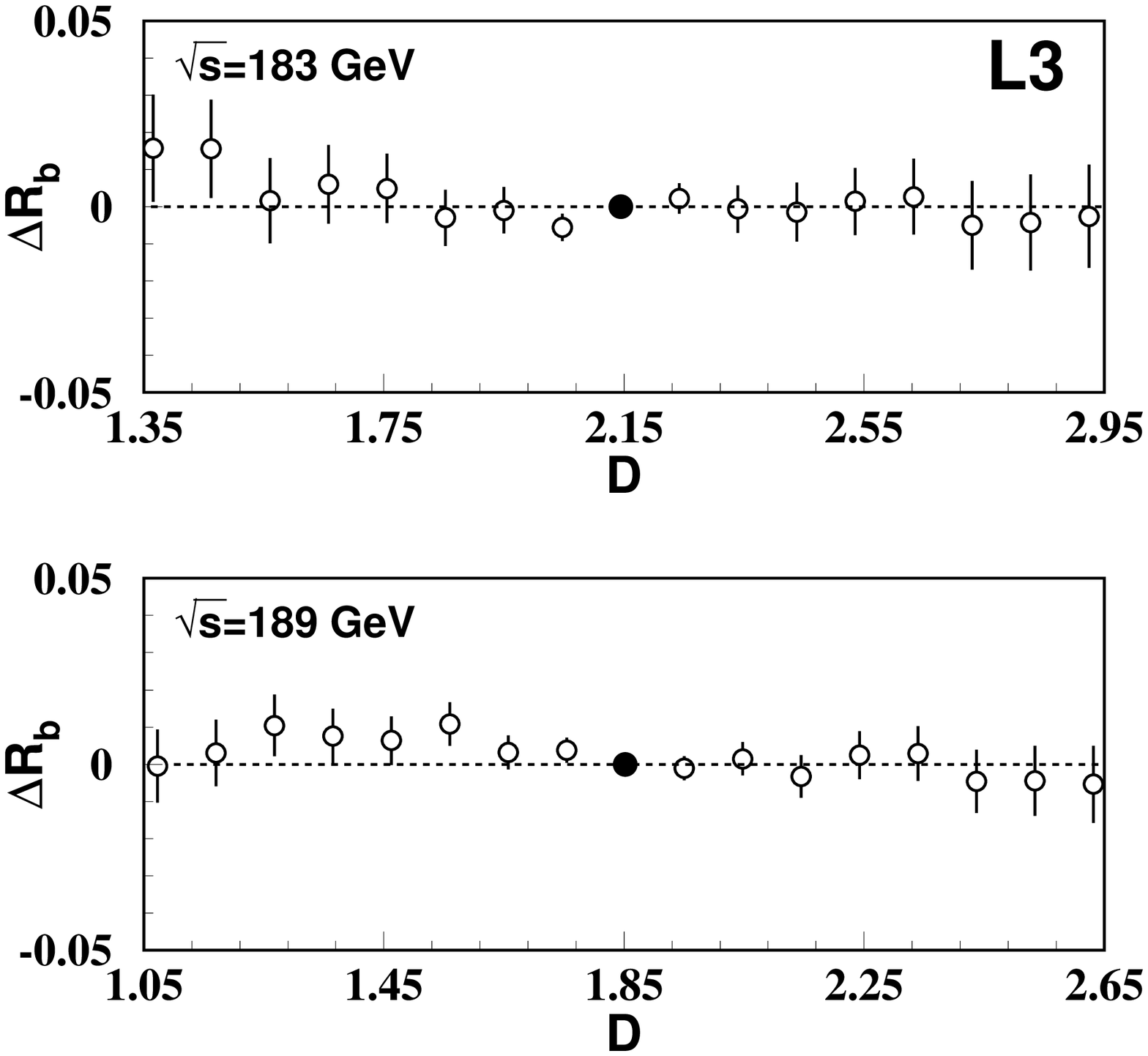, width=\textwidth}}
    \icaption{Variation of \RBB{} as a function of the cut on the
      b--tagging discriminant for the high statistics data samples at
      $\RS=183\GeV$ and 189 \GeV{}. The solid circles correspond to
      the measured values of \RBB.
  \label{fig:rbbvariation}}
  \end{center}
\end{figure}

The stability of the measurements has been checked by varying the cut
on the b--tagging discriminant. The results of this check
for the two high statistics samples at $\RS=183\GeV$ and 189 \GeV{}
are shown in Figure~\ref{fig:rbbvariation}. No statistically
significant deviation from the \RBB{} value obtained with the nominal
cut is observed.
The event--tag method used in this analysis has been cross--checked by
measuring \RBB{} using the calibration data collected each year at the
Z peak. The results are found to be in good agreement with our published
value \cite{RBBAFBL1L}. 

\section*{\boldmath Systematic uncertainties on \RBB}

The largest contribution to the systematic uncertainty on \RBB{} comes
from the description of the tracking resolution function in the Monte
Carlo, since it affects the estimation of the tagging efficiencies
involved in the \RBB{} measurement. In order to study this source of
systematic uncertainty, several Monte Carlo samples were
reconstructed, where the parameters used to describe the resolution
function have been varied according to their estimated uncertainty.
The differences between the values of $\varepsilon_{\rm b}$,
$\varepsilon_{\rm{c}}$ and $\varepsilon_{\rm{uds}}$ obtained from
these Monte Carlo samples and the nominal values shown in
Table~\ref{tab:rbbresults} have been propagated to obtain an estimation
of the systematic uncertainty due to tracking resolution effects.

Another source of systematic uncertainty comes from the description in the
Monte Carlo of the tracking efficiency and of the relative fraction of
tracks with a given SMD hit pattern. To study this effect, the tracking
efficiency has been varied in the Monte Carlo by $\pm 2\%$ and a
migration of tracks at the level of 1\% between the different classes
has been introduced. The systematic uncertainty coming from this source has
been obtained as above by propagating the variations on the flavour
tagging efficiencies. The systematic uncertainties discussed above
are added in quadrature to give the total systematic uncertainty due
to tracking effects.

The modelling of b and c quark fragmentation and decays introduces a
systematic uncertainty on \RBB{} since the tagging efficiency for b
and c quarks is obtained from Monte Carlo. 
Modelling uncertainties of the b--hadron properties
are estimated by varying the mean value of the b--hadron energy fraction 
$\langle x_E({\rm b})\rangle$, the charged decay multiplicity and the
average lifetime of b--hadrons.
For the estimation of $\varepsilon_{\rm{c}}$ an accurate knowledge of
production and decay properties of the c--hadrons is important, since the 
different species, $\Do$, $\Dp$, $\Ds$ and $\Lc$, have
lifetimes varying in the range of 0.2 to 1.1 ps \cite{PDG98}.
The variation of the parameters is performed following the recommendations 
of Reference \cite{LEPHF98}. A breakdown of the systematic uncertainties due to
the modelling of b and c--hadron physics is given in Table~\ref{tab:sysmod}. 

The error due to the finite Monte Carlo statistics has been estimated
by propagating the statistical errors on $\varepsilon_{\rm
  b}, \varepsilon_{\rm c}$ and $\varepsilon_{\rm uds}$.
The systematic uncertainty due to the selection of the $\EE\to\QQG$
event sample has been
estimated by varying the selection cuts and taking the corresponding
variation of \RBB{} as a systematic uncertainty. The error coming
from the non--\QQ{} and radiative background estimation was evaluated 
by varying the background fractions within their estimated uncertainties.

The various contributions to the systematic uncertainty on \RBB{} are
summarised in Table~\ref{tab:rbbsyst}.
 
\section*{\boldmath Measurement of \AFBB}

The forward--backward charge asymmetry of b quarks is measured by tagging
$\EE \to \BB$ events with electrons and muons with high momentum and
high transverse momentum with respect to the nearest jet. The leptons
are also used to identify the charge of the b quarks. The quark
direction is estimated using the event thrust axis. To enhance the b
purity, the b--tagging method previously described is
used.

\subsection*{Lepton identification and event selection}

Muons are identified in the muon chamber system. The muon tracks must
include track segments in at least two of the three layers of muon
chambers and must point to the interaction region.

Electrons are selected in the barrel region $|\cos\theta\,| < 0.7$.
They are identified by an energy deposit in the electromagnetic
calorimeter consistent with an electromagnetic shower matching with a
track in the central tracker. To reject misidentified hadrons,
the energy in the hadron calorimeter behind the electromagnetic shower
is required to be less than 6 \GeV{}.

The momentum of muon candidates is required to be greater than 6 \GeV{},
while that of the electron candidates must be greater than 3 \GeV{}. The
lepton transverse momentum, $p_{\rm{t}}$, is defined with respect to the
nearest jet, where the measured momentum of the lepton is not included
in the jet
reconstruction. If there is no jet with an energy greater than 6 \GeV{}
in the same hemisphere as the lepton, the $p_{\rm{t}}$ is calculated
relative to the thrust axis of the event. The transverse momentum
of the leptons must be greater than 0.7 \GeV{} to enhance
the \BB{} purity. In events containing more
than one lepton candidate, only the lepton with the highest
$p_{\rm{t}}$ is used in the asymmetry analysis. In order to further
reduce the background coming from radiative \QQ{} events and W pair
production, the angle $\alpha$ between the lepton and the event thrust
axis must satisfy the requirement $\rm{|\cos\alpha\,| > 0.95}$.

A cut in the ($p_{\rm{t}}$, $D$) plane is applied to enhance the b
purity, where $D$ is the discriminant variable used for the measurement
of \RBB. The cut is defined by:
\begin{eqnarray}
D  &>& -1.8\, p_{\rm{t}} + 2.4\quad\quad \mbox{(muons),}\nonumber\\
D  &>& -0.7\, p_{\rm{t}} + 1.6\quad\quad \mbox{(electrons).}
\end{eqnarray}
These cuts were chosen as a result of an optimisation procedure.

The number of prompt lepton candidates is 136, comprising 71 muon
candidates and 65 electron candidates in good agreement with the Monte
Carlo expectation of 139.4 events with 75.3 muons and 64.1 electrons.

Monte Carlo events are classified into eleven categories as listed in 
Table~\ref{tab:compo_final}. Leptons in b events not arising from the 
semi-leptonic decays of categories 1 to 5 are classed as ``b 
$\rightarrow$ fake lepton''.
This set of events consists of leptons from $\pi$ and $\mathrm{K}$
decays, Dalitz decays and photon conversions and of misidentified
hadrons. The 
Monte Carlo estimate of the sample composition is shown in
Table~\ref{tab:compo_final} together with the asymmetry contribution
of each event class.

\begin{figure}[ht]
\begin{center}
\begin{tabular}{cc}
\hspace{-1.5cm} 
  \epsfig{file=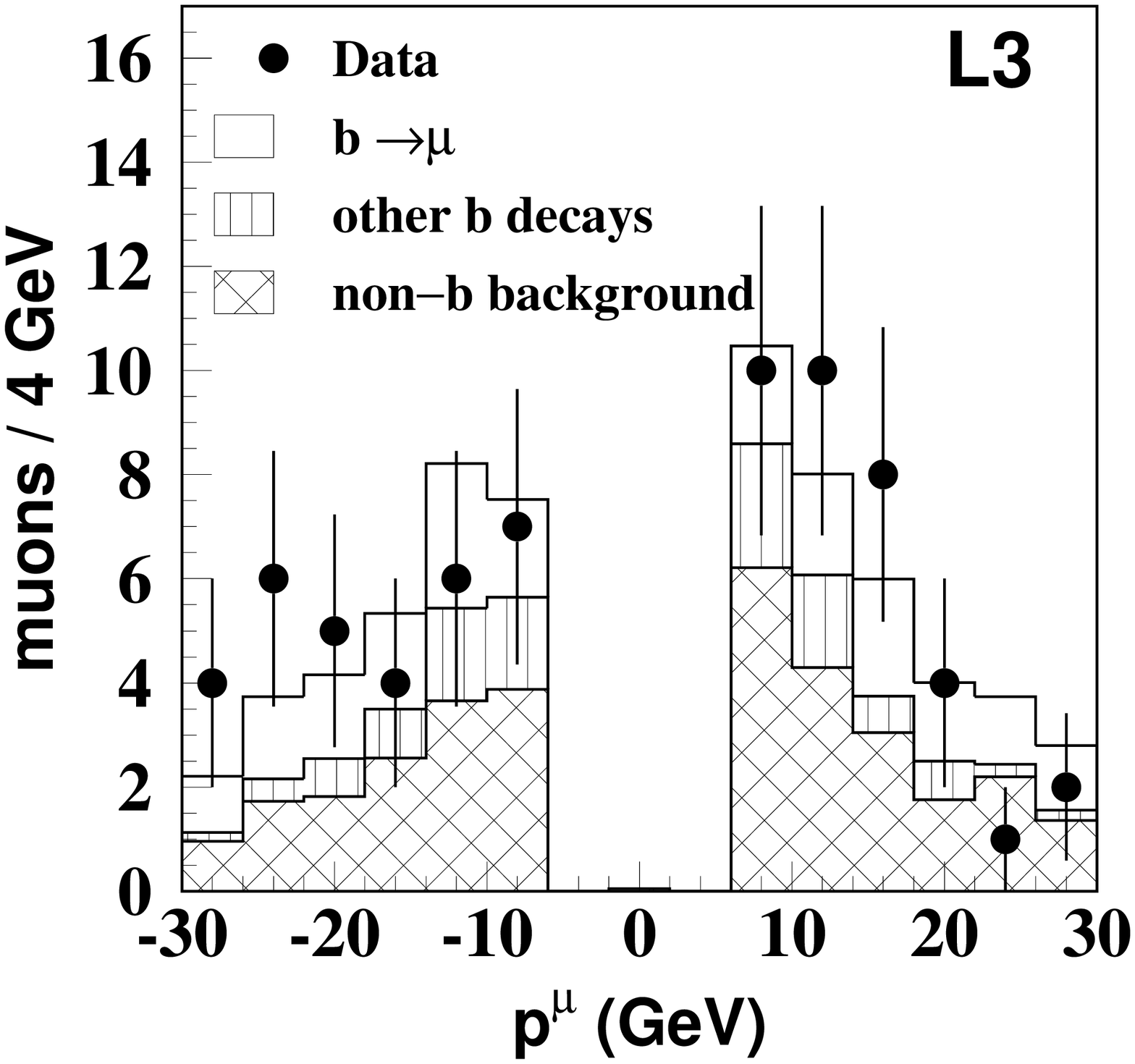,width=0.52\textwidth}
  \epsfig{file=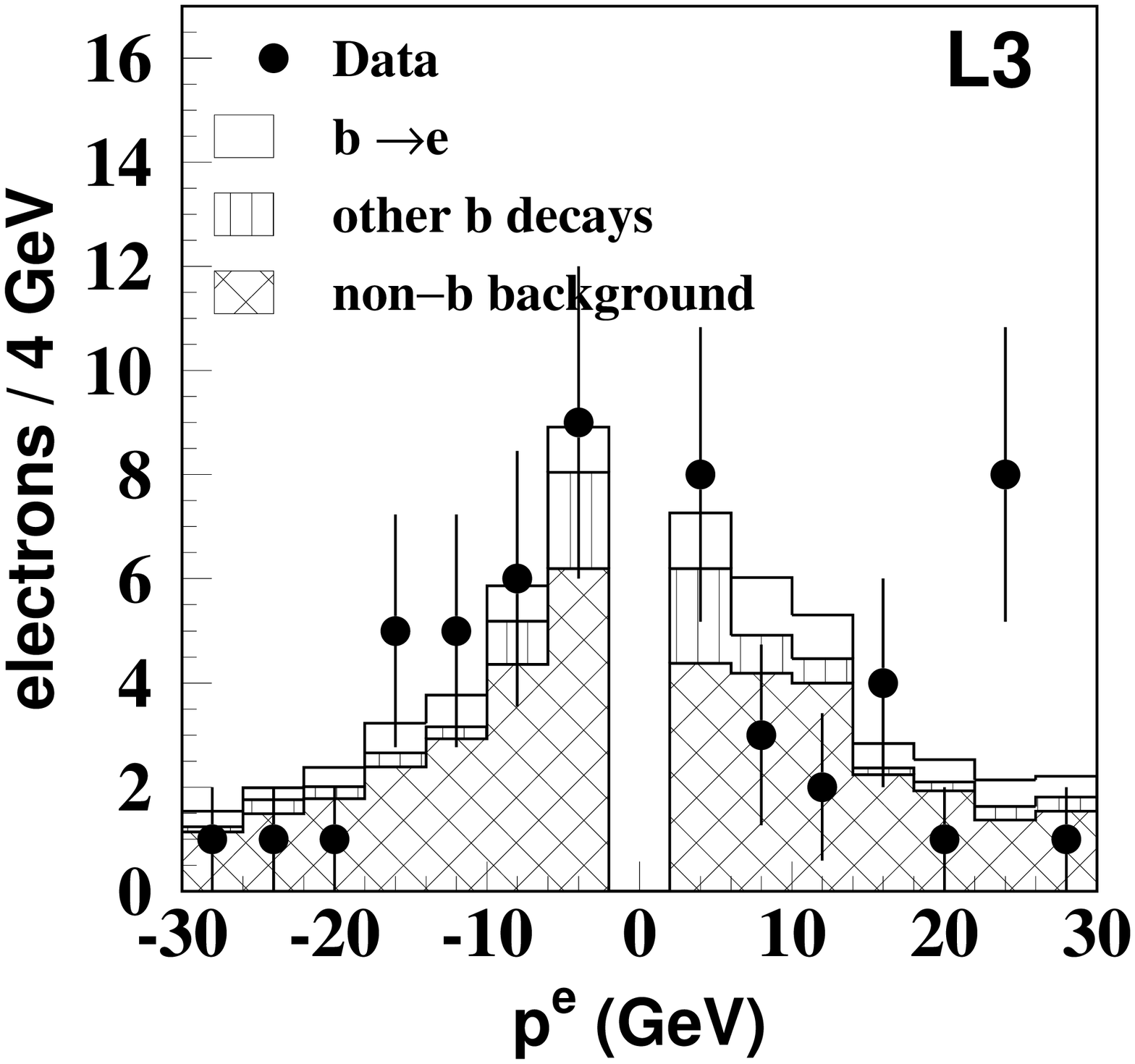,width=0.52\textwidth} \\
\hspace{-1.5cm} 
  \epsfig{file=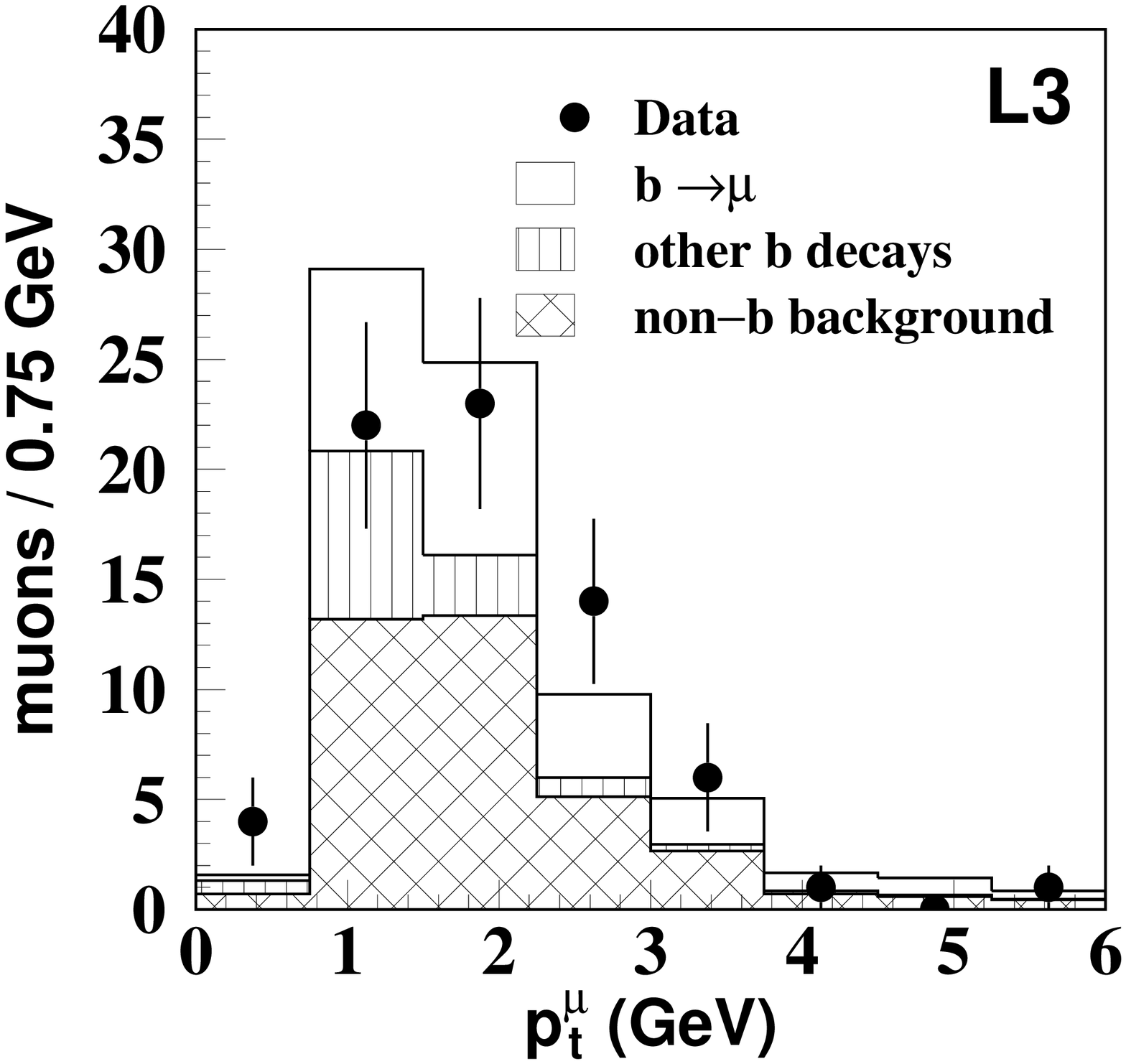,width=0.52\textwidth}
  \epsfig{file=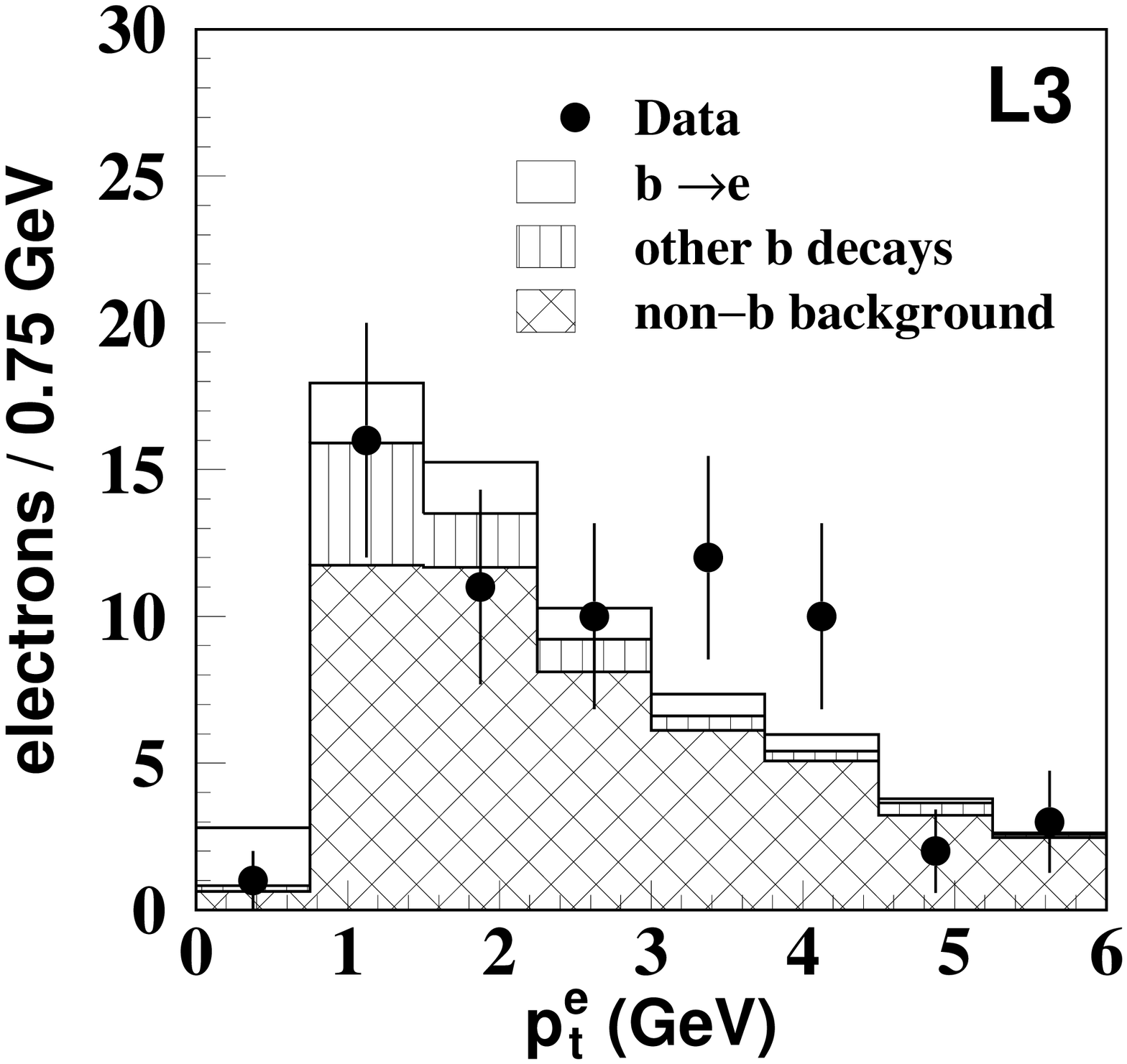,width=0.52\textwidth}
\end{tabular}
\icaption{Momentum and transverse momentum distributions for selected leptons.
  The histograms show the estimated composition of the sample
  according to Table~\ref{tab:compo_final}.
\label{fig:spectra}}
\end{center}
\end{figure}

The purities are $\mathrm{33.2\%}$ for the muon
sample and $\mathrm{16.8\%}$ for the electron sample. The 
efficiencies corresponding to these purities are
$\mathrm{21.2\%}$ and $\mathrm{7.6\%}$ respectively. 
The momentum and transverse
momentum spectra for the selected muon and electron candidates
together with the sample composition are shown in Figure~\ref{fig:spectra}.

\subsection*{Asymmetry determination}

\begin{figure}[ht]
\begin{center}
    \epsfig{file=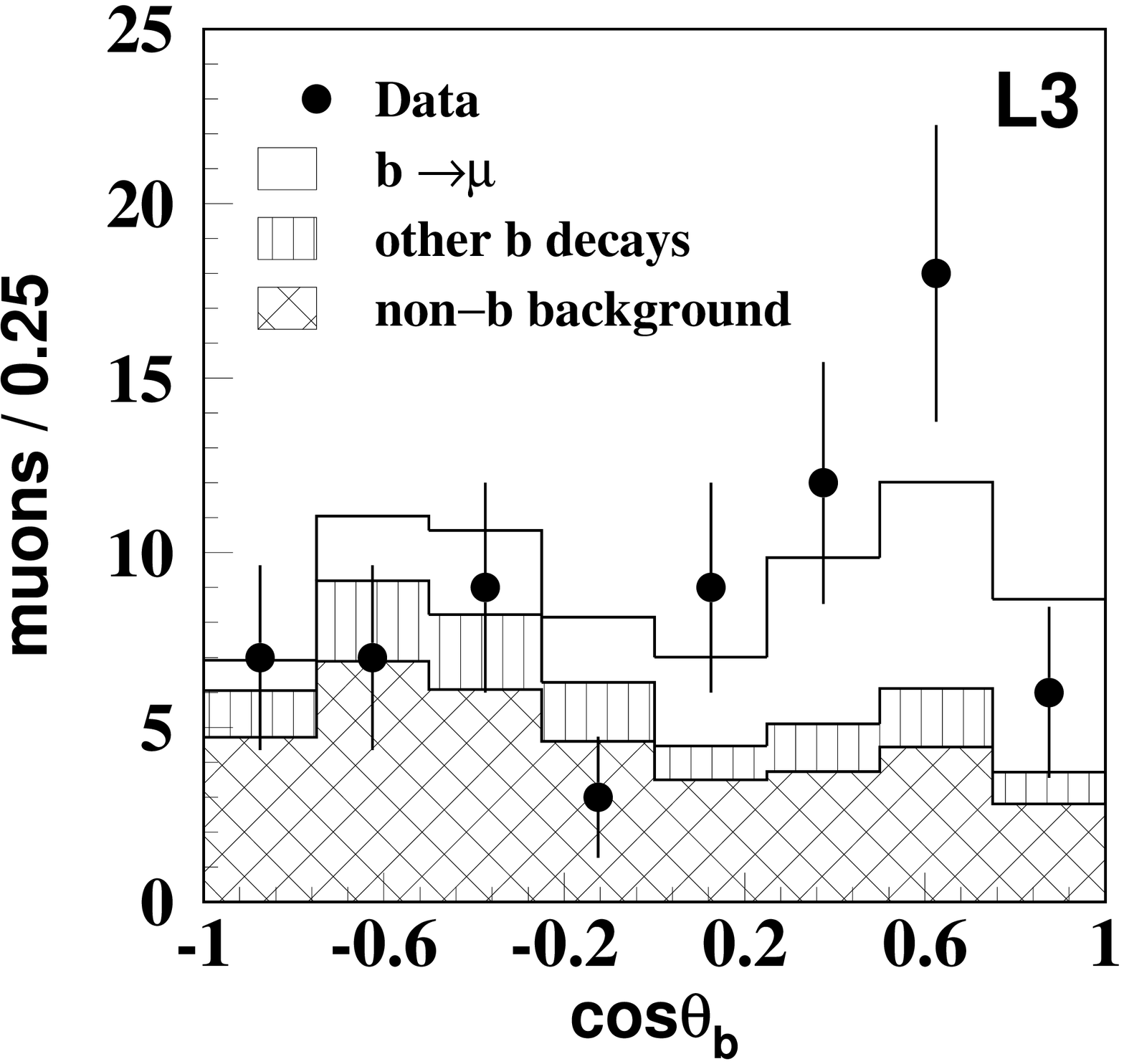,width=0.6\textwidth} 
    \epsfig{file=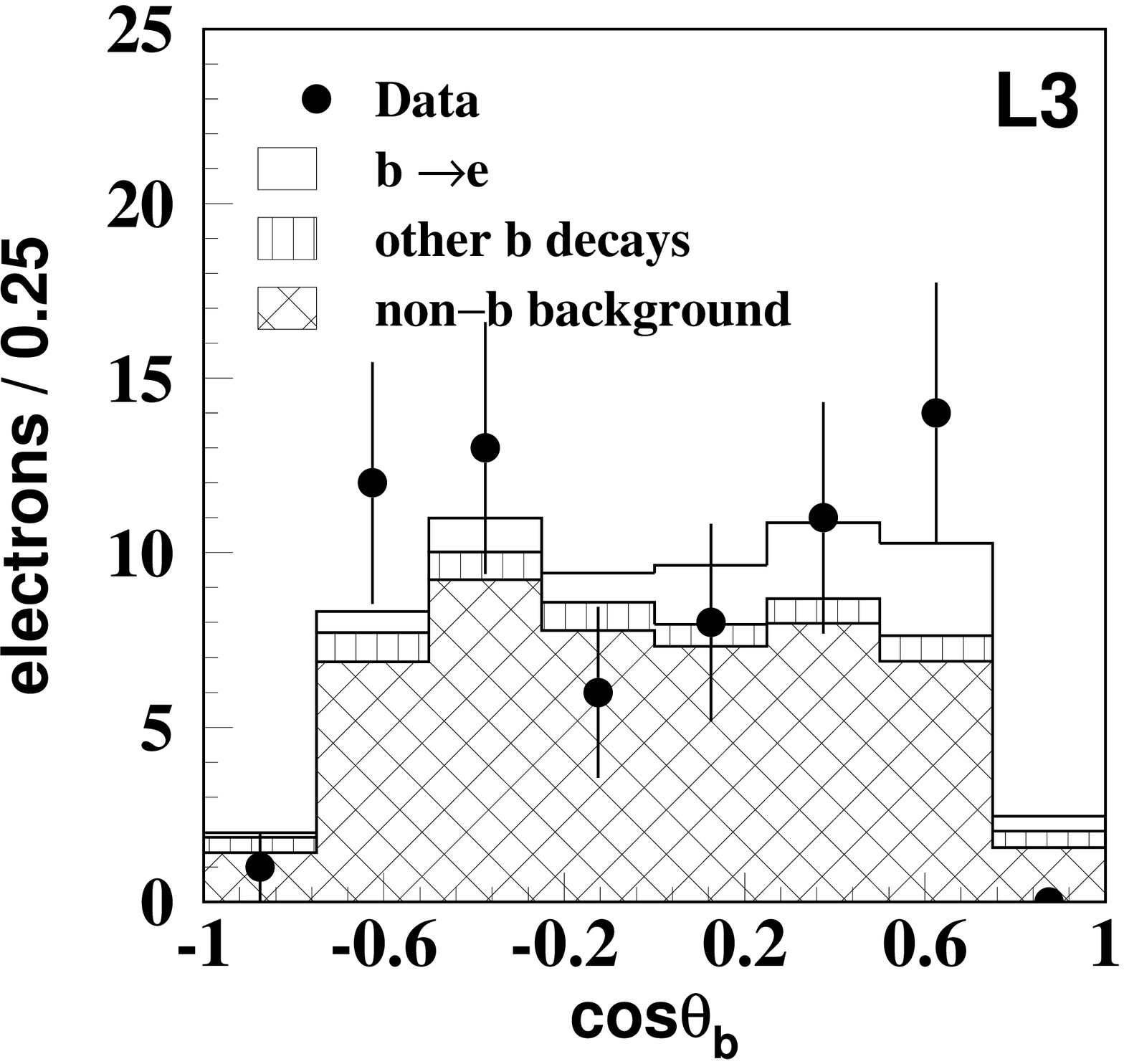,width=0.6\textwidth}
\icaption{Distribution of the scattering angle $\cos\theta_{\rm b} = -{\rm q
    \cos\theta_{\rm T}}$ for the muon and electron samples,
  respectively. The histograms represent the sample
  composition according to Table~\ref{tab:compo_final}.
\label{fig:asym}}
\end{center}
\end{figure}

The b quark scattering angle is obtained from the observable 
$\cos\theta_{\rm b}= -{q}\,\cos\theta_{\mathrm{T}}$, where the thrust direction
$\theta_{\mathrm{T}}$ is oriented into the hemisphere containing the
lepton and $q$ is the lepton charge.  The distribution of
$\cos\theta_{\rm b}$ for the selected muon and electron samples is shown in
Figure~\ref{fig:asym}.

Monte Carlo studies of the ``b 
$\rightarrow$ fake lepton'' class indicate a residual correlation between 
the quark charge and the observed charge, for both muons and electrons.
Variations of the generator level asymmetry reveal a linear correlation 
with the observed asymmetry for this class of events with 
$A_{\rm{bckg}}$=(0.16 $\pm$ 0.09)$\times A_{\rm b}$ for muons and 
(0.25 $\pm$ 0.08)$\times A_{\rm b}$ for electrons, where the
errors on the factors are statistical.  Variations by the
statistical errors on the factors are used to estimate 
systematic uncertainties arising from the $A_{\rm{bckg}}$
assignment.
A comparison with
results obtained using a constant $A_{\rm {bckg}}$ indicate differences
of 0.013 and 0.018 in the final measured asymmetries for muons and
electrons, respectively.

For each lepton $i$ in the data, the probability $f_k(i)$ to belong to 
each of the first six categories listed in Table~\ref{tab:compo_final} is 
determined from the number and type of Monte Carlo leptons found in the 
appropriate $\cos\theta_{\rm b}$ bin. 
The asymmetry is obtained by
applying a maximum-likelihood fit assuming no \BOBO mixing.
The likelihood function has the form:
\begin{equation}
  {\cal L} = \frac{\prod_{i=1}^{N_{\mathrm{data}}} \sum_{k=1}^{6}
  f_k(i) [ \frac{3}{8} ( 1 + \cos^2\theta_{i} ) + A_k\,
  \cos\theta_{i} ]}{\prod_{j=1}^{N_{\rm MC}^{\mathrm{non-b}}} (\sum_{k=1}^{6}
  f_k(j) [ \frac{3}{8} ( 1 + \cos^2\theta_{j} ) + A_k\,
  \cos\theta_{j}])^{W(j)}},
\label{eq:likelihood}
\end{equation}
where $A_k$ is the asymmetry contribution for the signal category $k$ 
and $W(j)$ represents the Monte Carlo to data normalisation relevant to the 
Monte Carlo lepton $j$, obtained from the ratio of the total number of
data and Monte Carlo leptons found in the corresponding
$\cos\theta_{\rm b}$ bin. The product over non--b Monte Carlo leptons in the 
denominator takes into account non--b background contributions in the data 
sample. 

The fitted asymmetries, corrected for the charge confusion ($2.0\pm 0.2\%$ for
electrons, and $0.1\pm 0.1\%$ for muons) and \BOBO mixing
($\chi_{\rm B}=0.1192 \pm 0.0068$ \cite{RBBAFBL1L,ASYMMETRY}), are
\begin{eqnarray}
  \AFBBm &=& 0.70 \pm 0.22~(stat.), \nonumber\\  
  \AFBBe &=& 0.39 \pm 0.34~(stat.).
\label{eq:afbres} 
\end{eqnarray}
\section*{\boldmath Systematic uncertainties on \AFBB}

The dependence of the measured asymmetry on other electroweak 
parameters \RBB{}, \RCC{} or \AFBC{} used in the Monte Carlo
modelling can be parametrised by :
\begin{equation} 
\Delta\AFBB = a(X)\Delta X
\label{eq:afbdep}
\end{equation}
where $\Delta X = X - X^{\rm SM}$ for $X=\RBB$, \RCC{}
or \AFBC{}. 
The values of $\RBB^{\rm SM}$, $\RCC^{\rm SM}$ and $\AFB^{\mathrm{c,SM}}$ used in the Monte 
Carlo are given in Table~\ref{tab:electro} together with the values of 
the coefficients $a(X)$.

Relevant fragmentation parameters, semileptonic decay models, and 
branching ratios are set to their measured values and their uncertainties 
are used to estimate the systematic uncertainty on the asymmetry.  The
parameter values and variations follow the recommendations developed
in Reference \cite{LEPHF98} and are listed in Table~\ref{tab:98lep} 
along with the corresponding systematic uncertainties on \AFBB.

Effects due to uncertainties in
the lepton momentum resolution are estimated by smearing the muon 
momentum by 1\% and the electron momentum by 5\%.
The uncertainty on the charge confusion correction is included as a
contribution to the systematic uncertainty. 

A possible bias on the measured asymmetry due to the non--b background
contribution is estimated by varying each background fraction by $\pm 5\%$ and 
the light quark asymmetry by $\pm 0.02$. 
The corresponding variations in \AFBB{} 
are taken as systematic uncertainties.

To study the Monte Carlo description of the $p_{\rm t}$ spectrum of the non--b
background, Monte Carlo events have been compared with high statistics 
background--enhanced data samples. Using these data
samples to constrain the shape of the $p_{\rm t}$ spectrum, Monte Carlo
reweighting factors for non--b background
are obtained in each $p_{\rm t}$ bin and used to modify the weights
$W(j)$ in Equation~\ref{eq:likelihood}.
Half of the change in the measured 
asymmetry caused by this reweighting procedure is
assigned as a systematic uncertainty from the background 
$p_{\rm t}$ description. 

The total systematic uncertainty on \AFBB{} is given in Table~\ref{tab:98lep}.

\section*{Results}

\begin{figure}[ht]
  \begin{center}
   \epsfig{file=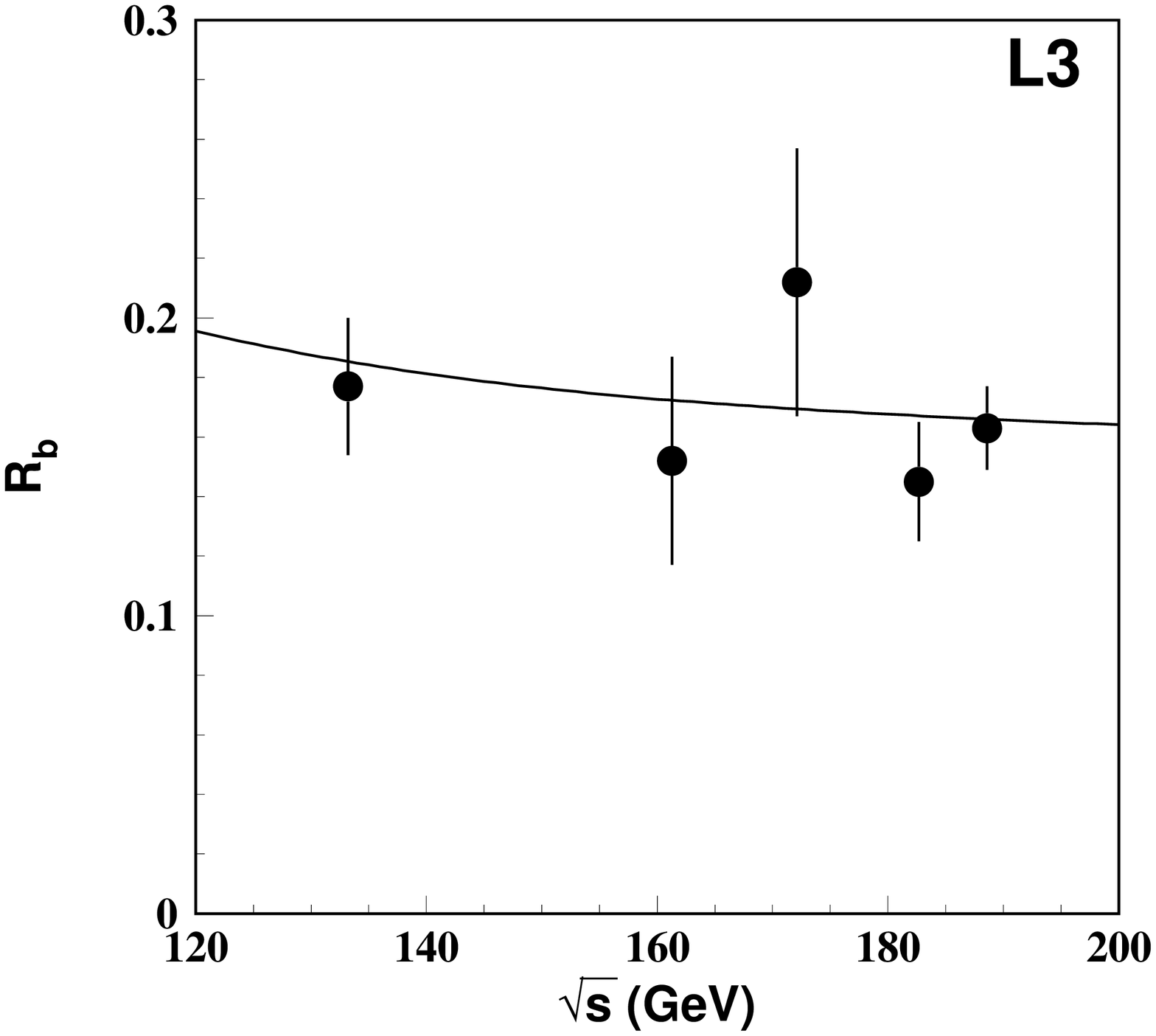,width=0.9\textwidth}
        \icaption[]{Measured values of \RBB{} as a function of
      centre--of--mass energy. The curve is the Standard Model
      prediction \cite{ZFITTER}. 
  \label{fig:rbbcomp}}
  \end{center}
\end{figure}

The measured values of \RBB{} are summarised in
Table~\ref{tab:results} with their statistical and systematic
errors. The measurements are compared to the Standard
Model prediction in Figure~\ref{fig:rbbcomp}.
At $\RS=189\GeV$ the measured value of \RBB{} is: 

$$ \RBB = 0.163 \pm 0.013~(stat.) \pm 0.005~(syst.) $$

\noindent to be compared with the Standard Model value \cite{ZFITTER} of 0.166.

The two measured values for \AFBBm{} and \AFBBe{} (Equation~\ref{eq:afbres})
are consistent and they are combined to give at $\RS=189\GeV$:

$$ \AFBB = 0.61 \pm 0.18~(stat.) \pm 0.09~(syst.)$$

\noindent to be compared with the Standard Model value \cite{ZFITTER} of 0.58.

%
%
\section*{Acknowledgements}

We wish to congratulate the CERN accelerator divisions for the
successful upgrade of the LEP machine and to express our gratitude for
its good performance. We acknowledge with appreciation the effort of
the engineers, technicians and support staff who have participated in
the construction and maintenance of this experiment.

%
%
\newpage
\typeout{   }     
\typeout{Using author list for paper 207 (Rbb) ONLY }
\typeout{Using author list for paper 207 (Rbb) ONLY }
\typeout{Using author list for paper 207 (Rbb) ONLY }
\typeout{Using author list for paper 207 (Rbb) ONLY }
\typeout{Using author list for paper 207 (Rbb) ONLY }
\typeout{Using author list for paper 207 (Rbb) ONLY }
\typeout{$Modified: Thu Apr  6 13:27:01 2000 by clare $}
\typeout{!!!!  This should only be used with document option a4p!!!!}
\typeout{   }
%
%
%
%
%
%

\newcount\tutecount  \tutecount=0
\def\tutenum#1{\global\advance\tutecount by 1 \xdef#1{\the\tutecount}}
\def\tute#1{$^{#1}$}
\tutenum\aachen            
\tutenum\nikhef            
\tutenum\mich              
\tutenum\lapp              
\tutenum\basel             
\tutenum\lsu               
\tutenum\beijing           
\tutenum\berlin            
\tutenum\bologna           
\tutenum\tata              
\tutenum\ne                
\tutenum\bucharest         
\tutenum\budapest          
\tutenum\mit               
\tutenum\debrecen          
\tutenum\florence          
\tutenum\cern              
\tutenum\wl                
\tutenum\geneva            
\tutenum\hefei             
\tutenum\seft              
\tutenum\lausanne          
\tutenum\lecce             
\tutenum\lyon              
\tutenum\madrid            
\tutenum\milan             
\tutenum\moscow            
\tutenum\naples            
\tutenum\cyprus            
\tutenum\nymegen           
\tutenum\caltech           
\tutenum\perugia           
\tutenum\cmu               
\tutenum\prince            
\tutenum\rome              
\tutenum\peters            
\tutenum\potenza           
\tutenum\salerno           
\tutenum\ucsd              
\tutenum\santiago          
\tutenum\sofia             
\tutenum\korea             
\tutenum\alabama           
\tutenum\utrecht           
\tutenum\purdue            
\tutenum\psinst            
\tutenum\zeuthen           
\tutenum\eth               
\tutenum\hamburg           
\tutenum\taiwan            
\tutenum\tsinghua          

{
\parskip=0pt
\noindent
{\bf The L3 Collaboration:}
\ifx\selectfont\undefined
 \baselineskip=10.8pt
 \baselineskip\baselinestretch\baselineskip
 \normalbaselineskip\baselineskip
 \ixpt
\else
 \fontsize{9}{10.8pt}\selectfont
\fi
\medskip
\tolerance=10000
\hbadness=5000
\raggedright
\hsize=162truemm\hoffset=0mm
\def\r{\rlap,}
\noindent

M.Acciarri\r\tute\milan\
P.Achard\r\tute\geneva\ 
O.Adriani\r\tute{\florence}\ 
M.Aguilar-Benitez\r\tute\madrid\ 
J.Alcaraz\r\tute\madrid\ 
G.Alemanni\r\tute\lausanne\
J.Allaby\r\tute\cern\
A.Aloisio\r\tute\naples\ 
M.G.Alviggi\r\tute\naples\
G.Ambrosi\r\tute\geneva\
% ...
\newpage
\begin{list}{A}{\itemsep=0pt plus 0pt minus 0pt\parsep=0pt plus 0pt minus 0pt
                \topsep=0pt plus 0pt minus 0pt}
\item[\aachen]
 I. Physikalisches Institut, RWTH, D-52056 Aachen, FRG$^{\S}$\\
 III. Physikalisches Institut, RWTH, D-52056 Aachen, FRG$^{\S}$
\item[\nikhef] National Institute for High Energy Physics, NIKHEF, 
     and University of Amsterdam, NL-1009 DB Amsterdam, The Netherlands
\item[\mich] University of Michigan, Ann Arbor, MI 48109, USA
\item[\lapp] Laboratoire d'Annecy-le-Vieux de Physique des Particules, 
     LAPP,IN2P3-CNRS, BP 110, F-74941 Annecy-le-Vieux CEDEX, France
\item[\basel] Institute of Physics, University of Basel, CH-4056 Basel,
     Switzerland
\item[\lsu] Louisiana State University, Baton Rouge, LA 70803, USA
\item[\beijing] Institute of High Energy Physics, IHEP, 
  100039 Beijing, China$^{\triangle}$ 
\item[\berlin] Humboldt University, D-10099 Berlin, FRG$^{\S}$
\item[\bologna] University of Bologna and INFN-Sezione di Bologna, 
     I-40126 Bologna, Italy
\item[\tata] Tata Institute of Fundamental Research, Bombay 400 005, India
\item[\ne] Northeastern University, Boston, MA 02115, USA
\item[\bucharest] Institute of Atomic Physics and University of Bucharest,
     R-76900 Bucharest, Romania
\item[\budapest] Central Research Institute for Physics of the 
     Hungarian Academy of Sciences, H-1525 Budapest 114, Hungary$^{\ddag}$
\item[\mit] Massachusetts Institute of Technology, Cambridge, MA 02139, USA
\item[\debrecen] KLTE-ATOMKI, H-4010 Debrecen, Hungary$^\P$
\item[\florence] INFN Sezione di Firenze and University of Florence, 
     I-50125 Florence, Italy
\item[\cern] European Laboratory for Particle Physics, CERN, 
     CH-1211 Geneva 23, Switzerland
\item[\wl] World Laboratory, FBLJA  Project, CH-1211 Geneva 23, Switzerland
\item[\geneva] University of Geneva, CH-1211 Geneva 4, Switzerland
\item[\hefei] Chinese University of Science and Technology, USTC,
      Hefei, Anhui 230 029, China$^{\triangle}$
\item[\seft] SEFT, Research Institute for High Energy Physics, P.O. Box 9,
      SF-00014 Helsinki, Finland
\item[\lausanne] University of Lausanne, CH-1015 Lausanne, Switzerland
\item[\lecce] INFN-Sezione di Lecce and Universit\'a Degli Studi di Lecce,
     I-73100 Lecce, Italy
\item[\lyon] Institut de Physique Nucl\'eaire de Lyon, 
     IN2P3-CNRS,Universit\'e Claude Bernard, 
     F-69622 Villeurbanne, France
\item[\madrid] Centro de Investigaciones Energ{\'e}ticas, 
     Medioambientales y Tecnolog{\'\i}cas, CIEMAT, E-28040 Madrid,
     Spain${\flat}$ 
\item[\milan] INFN-Sezione di Milano, I-20133 Milan, Italy
\item[\moscow] Institute of Theoretical and Experimental Physics, ITEP, 
     Moscow, Russia
\item[\naples] INFN-Sezione di Napoli and University of Naples, 
     I-80125 Naples, Italy
\item[\cyprus] Department of Natural Sciences, University of Cyprus,
     Nicosia, Cyprus
\item[\nymegen] University of Nijmegen and NIKHEF, 
     NL-6525 ED Nijmegen, The Netherlands
\item[\caltech] California Institute of Technology, Pasadena, CA 91125, USA
\item[\perugia] INFN-Sezione di Perugia and Universit\'a Degli 
     Studi di Perugia, I-06100 Perugia, Italy   
\item[\cmu] Carnegie Mellon University, Pittsburgh, PA 15213, USA
\item[\prince] Princeton University, Princeton, NJ 08544, USA
\item[\rome] INFN-Sezione di Roma and University of Rome, ``La Sapienza",
     I-00185 Rome, Italy
\item[\peters] Nuclear Physics Institute, St. Petersburg, Russia
\item[\potenza] INFN-Sezione di Napoli and University of Potenza, 
     I-85100 Potenza, Italy
\item[\salerno] University and INFN, Salerno, I-84100 Salerno, Italy
\item[\ucsd] University of California, San Diego, CA 92093, USA
\item[\santiago] Dept. de Fisica de Particulas Elementales, Univ. de Santiago,
     E-15706 Santiago de Compostela, Spain
\item[\sofia] Bulgarian Academy of Sciences, Central Lab.~of 
     Mechatronics and Instrumentation, BU-1113 Sofia, Bulgaria
\item[\korea]  Laboratory of High Energy Physics, 
     Kyungpook National University, 702-701 Taegu, Republic of Korea
\item[\alabama] University of Alabama, Tuscaloosa, AL 35486, USA
\item[\utrecht] Utrecht University and NIKHEF, NL-3584 CB Utrecht, 
     The Netherlands
\item[\purdue] Purdue University, West Lafayette, IN 47907, USA
\item[\psinst] Paul Scherrer Institut, PSI, CH-5232 Villigen, Switzerland
\item[\zeuthen] DESY, D-15738 Zeuthen, 
     FRG
\item[\eth] Eidgen\"ossische Technische Hochschule, ETH Z\"urich,
     CH-8093 Z\"urich, Switzerland
\item[\hamburg] University of Hamburg, D-22761 Hamburg, FRG
\item[\taiwan] National Central University, Chung-Li, Taiwan, China
\item[\tsinghua] Department of Physics, National Tsing Hua University,
      Taiwan, China
\item[\S]  Supported by the German Bundesministerium 
        f\"ur Bildung, Wissenschaft, Forschung und Technologie
\item[\ddag] Supported by the Hungarian OTKA fund under contract
numbers T019181, F023259 and T024011.
\item[\P] Also supported by the Hungarian OTKA fund under contract
  numbers T22238 and T026178.
\item[$\flat$] Supported also by the Comisi\'on Interministerial de Ciencia y 
        Tecnolog{\'\i}a.
\item[$\sharp$] Also supported by CONICET and Universidad Nacional de La Plata,
        CC 67, 1900 La Plata, Argentina.
\item[$\diamondsuit$] Also supported by Panjab University, Chandigarh-160014, 
        India.
\item[$\triangle$] Supported by the National Natural Science
  Foundation of China.
\item[\dag] Deceased.
\end{list}
}
\vfill


\newpage
%

\newpage


%
%
%

\begin{table}[tp]
  \begin{center}
    \begin{tabular}{|c|c|c|c|c|c|} \hline
      \RS{} (\GeV{}) & 133.2 & 161.3 & 172.1 & 182.7 & 188.6\\\hline
      \IL{} (\IVPB) & 12.01 & 10.95 & 10.25 & 55.65 & 176.3
      \\\hline
    \end{tabular}
  \end{center}
  \icaption{Summary of centre--of--mass energies and integrated
  luminosities considered in this analysis. 
  \label{tab:enelumi}}
\end{table}

%
%
\begin{table}[tp]
  \begin{center}
    \begin{tabular}{|c|r|r|r|r|r|} \hline
 \RS{} (\GeV{}) 
&\multicolumn{1}{|c|}{133} 
&\multicolumn{1}{|c|}{161} 
&\multicolumn{1}{|c|}{172} 
&\multicolumn{1}{|c|}{183} 
&\multicolumn{1}{|c|}{189}\\\hline
      $\RCC^{\rm SM}$ & 0.221 & 0.242 & 0.246 & 0.250 & 0.252 \\\hline
    $a(\RCC)$  & $-0.113$ & $-0.118$ & $-0.106$ & $-0.099$ & $-0.100$ \\\hline
    \end{tabular}
  \end{center}
  \icaption[]{Values of $\RCC^{\rm SM}$ used for the measurement of \RBB{} and
  coefficients $a(\RCC)$ giving the dependence of the measured \RBB{} on
  \RCC{} (Equation~\ref{eq:rccdep}). 
  \label{tab:rccdep}}
\end{table}

%
%

\begin{table}[bp]
  \begin{center}
    \begin{tabular}{|c|r|r|r|r|r|r|} \hline
\RS{} (\GeV{}) 
&\multicolumn{1}{|c|}{$N^{\rm{obs}}$}
&\multicolumn{1}{|c|}{$N\rm{_{t}^{obs}}$} 
&\multicolumn{1}{|c|}{$N^{\rm{bkg}}$}
&\multicolumn{1}{|c|}{$N\rm{^{bkg}_t}$} 
&\multicolumn{1}{|c|}{tagging} 
&\multicolumn{1}{|c|}{\RBB}\\
      &&&&&\multicolumn{1}{|c|}{efficiencies}& \\
      \hline\hline 
      &&&&&$\varepsilon_{\rm b}    =$0.676$\pm$0.007&  \\
      133 & 850 & 134 & $138.5\pm 2.0$ &$23.8\pm 0.4$ &
      $\varepsilon_{\rm c}=$0.097$\pm$0.004 & 0.177$\pm$0.023\\ 
      &&&&&$\varepsilon_{\rm uds}=$0.023$\pm$0.001&  \\   
      \hline                                           
      &&&&&$\varepsilon_{\rm b}    =$0.645$\pm$0.006&  \\   
      161 & 319 & 45 & $28.2\pm 0.4$ &$3.9\pm 0.2$  &  
      $\varepsilon_{\rm c}=$0.103$\pm$0.003 & 0.152$\pm$0.035\\
      &&&&&$\varepsilon_{\rm uds}=$0.031$\pm$0.001&  \\   
      \hline                                           
      &&&&&$\varepsilon_{\rm b}    =$0.619$\pm$0.006&  \\   
      172 & 266 & 43 & $38.0\pm 0.4$ & $3.7\pm 0.1$ &  
      $\varepsilon_{\rm c}=$0.095$\pm$0.003 & 0.212$\pm$0.045\\
      &&&&&$\varepsilon_{\rm uds}=$0.032$\pm$0.001&  \\   
      \hline                                           
      &&&&&$\varepsilon_{\rm b}    =$0.621$\pm$0.004&  \\   
      183 & 1172 & 147 & $185.8\pm 0.9$ &$16.6\pm 0.3$ & 
      $\varepsilon_{\rm c}=$0.091$\pm$0.002 & 0.145$\pm$0.020\\
      &&&&&$\varepsilon_{\rm uds}=$0.033$\pm$0.001&  \\   
      \hline                                           
      &&&&&$\varepsilon_{\rm b}    =$0.600$\pm$0.002&  \\   
      189 & 3462 & 486 & $524.3\pm 2.3$ & $50.6\pm 0.6$ &
      $\varepsilon_{\rm c}=$0.099$\pm$0.001 & 0.163$\pm$0.013\\
      &&&&&$\varepsilon_{\rm uds}=$0.044$\pm$0.001&  \\
      \hline
    \end{tabular}
  \end{center}
  \icaption{Number of selected ($N^{\rm{obs}}$) and tagged
    ($N\rm{_{t}^{obs}}$) events, background
    contamination both in the total ($N^{\rm{bkg}}$) and in the tagged
    sample ($N\rm{^{bkg}_t}$) and tagging
    efficiencies for each centre--of--mass energy. The errors on the
    tagging efficiencies are statistical. In the last column the
    measured values of \RBB{} are listed. The errors are statistical only.
    \label{tab:rbbresults}}
\end{table}

\clearpage

%
%

\begin{table}[tp]
  \begin{center}
    \begin{tabular}{|l|r|r|r|r|r|}\hline
      \multicolumn{1}{|c|}{Source} &
      \multicolumn{5}{|c|}{$\Delta\RBB$}\\\cline{2-6}
      \multicolumn{1}{|c|}{}& 133 \GeV{} & 161 \GeV{} &
      172 \GeV{} & 183 \GeV{} & 189 \GeV{}
      \\\hline\hline 
b fragmentation     & 0.0015 & 0.0014 & 0.0020 & 0.0014 & 0.0016 \\\hline
b lifetime          & 0.0003 & 0.0003 & 0.0004 & 0.0003 & 0.0003 \\\hline
b decay multiplicity& 0.0010 & 0.0009 & 0.0013 & 0.0009 & 0.0011 \\\hline
c modelling         & 0.0005 & 0.0006 & 0.0006 & 0.0006 & 0.0006 \\\hline\hline
Total b, c physics modelling & 0.0019 & 0.0018 & 0.0025 & 0.0018 & 0.0021 \\\hline 
    \end{tabular}
  \end{center}
  \icaption{Summary of systematic uncertainties on \RBB{} due to b and
  c--hadron physics modelling.
  \label{tab:sysmod}}

\end{table}

%
%

\begin{table}[tp]
  \begin{center}
    \begin{tabular}{|l|r|r|r|r|r|}\hline
      \multicolumn{1}{|c|}{Source} &
      \multicolumn{5}{|c|}{$\Delta\RBB$}\\\cline{2-6}
      \multicolumn{1}{|c|}{}& 133 \GeV{} & 161 \GeV{} &
      172 \GeV{} & 183 \GeV{} & 189 \GeV{}
      \\\hline\hline 
Tracking effects      & 0.004 & 0.004 & 0.005 & 0.003 & 0.004 \\\hline 
b, c physics modelling & 0.002 & 0.002 & 0.003 & 0.002 & 0.002 \\\hline 
MC statistics         & 0.003 & 0.002 & 0.003 & 0.001 & 0.001 \\\hline
Event selection       & 0.002 & 0.002 & 0.002 & 0.002 & 0.002 \\\hline
Background            & 0.001 & 0.001 & 0.001 & 0.001 & 0.001 \\\hline\hline
Total systematic uncertainty& 0.005 & 0.005 & 0.007 & 0.004 & 0.005 \\\hline 
    \end{tabular}
  \end{center}
  \icaption{Summary of systematic uncertainties on \RBB.
  \label{tab:rbbsyst}}

\end{table}

\clearpage

%
%
%
%

\begin{table}[btp]
 \begin{center}
  \begin{tabular}{|l|r|r|c|}\hline
Category & Muons (\%) & Electrons (\%) & $A_{k}$\\ 
\hline \hline 
$\rm {b \to l}$ & 33.20 $\pm$ 0.40  & 16.80 $\pm$ 0.30 & \pho $A_{\rm{b}}$ \\ 
\hline  
$\rm {b\to c\to l}$  & 8.20 $\pm$ 0.20 & 1.45 $\pm$ 0.10 & $-A_{\rm{b}}$ \\ 
\hline
$\rm {b\to \bar{c}\to l}$ & 1.80 $\pm$ 0.10 & 0.45 $\pm$ 0.05 & \pho $A_{\rm{b}}$ \\
\hline
$\rm {b\to J/\psi\to l}$ & 0.40 $\pm$ 0.05 & \multicolumn{1}{c|}{$-$} & 0 \\ 
\hline 
$\rm {b\to\tau\to l}$  & 1.10 $\pm$ 0.09 & 0.70 $\pm$ 0.06 & \pho $A_{\rm{b}}$ \\ 
\hline 
$\rm {b\to fake~l}$ & 5.70 $\pm$ 0.20 & 12.90 $\pm$ 0.30 & $A_{\rm{bckg}}$\\
\hline
$\rm {c\to l}$ & 14.70 $\pm$ 0.35  & 2.10 $\pm$ 0.10 &  $-$\\ 
\hline 
$\rm {c\to fake~l}$ & 5.50 $\pm$ 0.40 & 12.70 $\pm$ 0.45 & $-$\\
\hline
$\rm {u, d, s\to l, fake~l}$ & 21.10 $\pm$ 0.25 & 37.20 $\pm$ 0.30 &  $-$\\
\hline
rad. background & 3.80 $\pm$ 0.20  & 4.40 $\pm$ 0.10   &  $-$\\ 
\hline 
WW background  & 4.50  $\pm$ 0.20  & 11.30 $\pm$ 0.45  &  $-$\\ 
\hline
  \end{tabular}
 \end{center}
 \icaption{Fractions of events in the various categories described
   in the text, estimated from Monte Carlo. The corresponding asymmetry
   contributions used in the likelihood fit are shown in the last
   column. 
\label{tab:compo_final}}
\end{table}
\clearpage
%
%
\begin{table}[bp]
\begin{center}
 \begin{tabular}{|l|c|}\hline 
Standard Model value & $a(X)$ \\ \hline
$\RBB^{\rm SM} = 0.166$   & $-0.81$\\
$\RCC^{\rm SM} = 0.252$   & $+1.44$\\
$\AFB^{\mathrm{c,SM}} = 0.62$  & $+0.47$\\\hline
\end{tabular}
\end{center}
 \icaption[]{Values of the \SM{} parameters $\RBB^{\rm SM}$,
 $\RCC^{\rm SM}$ and
 $\AFB^{\mathrm{c,SM}}$ used in the \AFBB{} analysis and coefficients
 $a(X)$ giving the dependence of \AFBB{} on them (Equation~\ref{eq:afbdep}).
\label{tab:electro}}
\end{table}

%
%

\begin{table}[bp]
\begin{center}
 \begin{tabular}{|lccc|}\hline 
Contribution & Value &Variation & $\Delta \AFBB$\\ \hline


$x_{E}\mathrm{(b)}$          &  0.702   & $\pm$ 0.008 & $\mp$  0.004  \\
$x_{E}\mathrm{(c)}$          &  0.484   & $\pm$ 0.008 & $\pm$  0.006  \\
$\rm Br(b\to l)$             &  0.105   & $\pm$ 0.005 & $\mp$  0.007  \\
$\rm Br(b\to c\to l)$        &  0.080   & $\pm$ 0.005 & $\pm$  0.004  \\
$\rm Br(b\to\bar{c}\to l)$   &  0.013   & $\pm$ 0.005 & $\mp$  0.001  \\
$\rm Br(b\to\tau\to l)$      &  0.005   & $\pm$ 0.001 & $\mp$  0.002  \\
$\rm Br(c\to l)$             &  0.098   & $\pm$ 0.005 & $\pm$  0.013  \\
$\rm Br(b\to J/\psi\to l)$   &  0.001   & $\pm$ 0.001 & $\pm$  0.002  \\
 \multicolumn{3}{|l}{Fragmentation and Branching Ratios} & 0.017 \\\hline\hline

 $\rm b \to l $ model &  &             & $\mp$      0.005 \\
 $\rm c \to l $ model &  &             & $\pm$      0.025 \\
 $\rm b \to D $ model &  &             & $\pm$      0.003 \\
 \multicolumn{3}{|l}{Decay Models} &       0.026     \\\hline\hline


background fractions       &       & $\pm 5\%$ & $\pm$       0.024  \\
u, d, s background asymmetry   & 0.00  & $\pm 0.02$ & $\mp$       0.020  \\ 
$A_{\rm bckg}$ asymmetry   &       & $\pm 0.08$ & $\mp$       0.004 \\
 \multicolumn{3}{|l}{ Background effects} &  0.031     \\\hline\hline

$p_{\rm t}$ background description  &       &        & $\pm$ 0.073 \\ 
lepton momentum smearing    &       &      & $\pm$       0.005  \\
charge confusion correction &       &           & $\mp$       0.002  \\
mixing dependence           &  0.1192     & 0.0068 & $\pm$ 0.010 \\
 \multicolumn{3}{|l}{Detector effects} &       0.074 \\\hline\hline
 \multicolumn{3}{|l}{ Total systematic uncertainty} &       0.087\\\hline
 \end{tabular}
\end{center}
 \icaption{Systematic uncertainties on \AFBB. The 
values and variations for the charge confusion correction and momentum 
smearing are given in the text.
\label{tab:98lep}}
\end{table}
\clearpage

%
%

\begin{table}[btp]
  \begin{center}
    \begin{tabular}{|l|c|c|} 
      \hline
      \RS{} (\GeV{}) & \RBB  & SM Expectation\\ 
\hline\hline
      133.2  &$0.177\pm 0.023\pm 0.005$ & 0.185 \\ \hline
      161.3  &$0.152\pm 0.035\pm 0.005$ & 0.172 \\ \hline
      172.1  &$0.212\pm 0.045\pm 0.007$ & 0.169 \\ \hline
      182.7  &$0.145\pm 0.020\pm 0.004$ & 0.167 \\ \hline
      188.6  &$0.163\pm 0.013\pm 0.005$ & 0.166 \\ \hline
    \end{tabular}
  \end{center}
  \icaption[]{Summary of results on \RBB. The first error is
    statistical, and the second is systematic. The values predicted by the
    \SM \cite{ZFITTER} are shown in the last column.
  \label{tab:results}} 
\end{table}

\clearpage



\end{document}